\documentclass[
reprint, twocolumn
 ,secnumarabic%
,amssymb,
nobibnotes,
nofootinbib,
aps, prl,
showpacs,showkeys]{revtex4}
\usepackage{docs}%
\usepackage{bm}%
\usepackage{graphicx}
\usepackage{hyperref}
\usepackage{url}

\expandafter
\ifx
\csname package@font\endcsname\relax\else
\expandafter
\expandafter
\expandafter
\usepackage
\expandafter
\expandafter
\expandafter{\csname package@font\endcsname}%
\fi



\begin{document}

\title{Two-Photon Exchange Effect Studied with Neural Networks}
\author{Krzysztof M. Graczyk}

\email{kgraczyk@ift.uni.wroc.pl}
\affiliation{Institute of Theoretical Physics, University of Wroc\l aw, pl. M. Borna 9, 50-204, Wroc\l aw, Poland}
\date{\today}%

\begin{abstract}
An approach to the extraction of the  two-photon exchange (TPE) correction from elastic $ep$ scattering data is presented. The cross section, polarization transfer (PT), and charge asymmetry data are considered. It is assumed that the TPE correction to the PT data is negligible. The form factors and TPE correcting term are given by  one multidimensional function approximated by the feed forward  neural network (NN). To find a model-independent approximation  the Bayesian framework for the NNs is adapted.    A large number of different parametrizations is considered. The most optimal model is indicated by the Bayesian algorithm. The obtained fit of the TPE correction behaves  linearly in $\epsilon$ but it has a nontrivial  $Q^2$ dependence. A strong dependence of the TPE fit on the choice of parametrization is observed.
\end{abstract}

\maketitle



\section{Introduction}

The study of elastic $ep$ scattering provides an opportunity to explore the structure of the proton. From $ep$ cross section data the  magnetic ($G_M$) and electric ($G_E$) proton form factors (FFs)  are obtained via longitudinal-transverse (LT) separation \cite{Perdrisat:2006hj,Arrington:2006zm}. In the data analysis, it is convenient to consider the reduced cross section, which in the one-photon exchange (OPE) approximation, reads as
\begin{eqnarray}
\label{sigma_R}
\sigma_{1\gamma,R}(Q^2, \epsilon) &=& \tau G_M^2(Q^2) + \epsilon G_E^2(Q^2), \\
\tau  = Q^2/4M^2, \; & &
\epsilon =\left(1 + 2 (1 + \tau )\tan^2(\theta/2)\right)^{-1}, \nonumber
\end{eqnarray}
where $Q^2$ and $\theta$ are the four-momentum transfer squared and  scattering angle respectively.
\begin{figure}
\centering{
\includegraphics[scale=0.5]{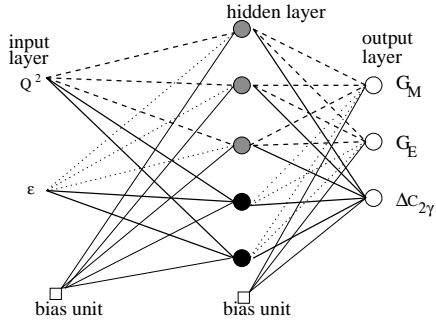}}
\caption{ The 2-(3-2)-3 type network, two input units, one layer of hidden units, and three output units. The FF sector (gray filled units and dashed connections) in contrast to the TPE sector (black units and solid connections) is connected only with $Q^2$.  Dotted lines denote  the switched-off connections. Solid and dashed lines represent the weight parameters. The bias unit is not connected with the units from the previous layer, and its signal is equal to 1, which means that $f_{act}^{bias}=1$.
\label{Fig_sieci} }
\end{figure}

The FF ratio,
\begin{equation}
\mathcal{R}_{1\gamma}(Q^2) = \mu_p \frac{G_E(Q^2)}{ G_M(Q^2)}
\end{equation}
($\mu_p$, the magnetic moment of the proton) can be   extracted from the so-called polarization transfer (PT) measurements \cite{Ron:2011rdand_Puckett:2011xg}. It turns out that  the systematic discrepancy between the FF ratio data obtained via the LT separation and the PT  measurements exists. It seems that taking into account the two-photon exchange (TPE) correction, the one which is not included in the  classical treatment of the radiative corrections \cite{Mo:1968cg}, cancels this discrepancy \cite{Guichon:2003qm,Blunden:2003sp}. Moreover, it is claimed that TPE correction to the $\mu_p G_E/G_M$ ratio, extracted from the PT measurements, is negligible \cite{Guichon:2003qm,Meziane:2010xc}. However, taking into account the  TPE contribution, in the LT separation, affects   significantly the extracted values of the proton FFs. The reduced cross section is modified by the TPE correction $\Delta C_{2\gamma}(Q^2, \epsilon)$, namely,
\begin{equation}
\label{sigma_r_2gamma}
\sigma_{1\gamma+2\gamma, R}(Q^2, \epsilon) =
\sigma_{1\gamma,R}(Q^2,\epsilon) + \Delta C_{2\gamma}(Q^2, \epsilon).
\end{equation}

The TPE effect has been  studied extensively during the past few years, for reviews and references, see Refs. \cite{Carlson:2007sp,Arrington:2011dn}.
The recent studies can be found in Refs. \cite{Meziane:2010xc,Qattan:2011zz,Borisyuk:2010ep,Guttmann:2010au}.

The dominant part of $\Delta C_{2\gamma}(Q^2, \epsilon)$  is given by the interference between the OPE and the TPE  amplitudes.
Hence, for the $e^+p$ scattering,
\begin{equation}
\Delta C_{2\gamma}(e^+p) = -\Delta C_{2\gamma}(e^-p).
 \end{equation}
 Therefore, the magnitude  of  the TPE term can be evaluated by measuring the ratio of the $e^+p$ to $e^- p$ elastic cross sections \cite{positron data,Yount62},
\begin{equation}
\mathcal{R}_{\pm}(Q^2,\epsilon) =
1 - \frac{2\Delta C_{2\gamma}(Q^2,\epsilon)}{\sigma_{1\gamma+2\gamma,R}(Q^2,\epsilon)}.
\end{equation}
A deviation of this function from unity  indicates the importance of the TPE effect \cite{Arrington:2003ck}.

A direct prediction of the proton FFs and TPE correction is a difficult task. One has to deal with the problems  of quantum chromodynamics in the non-perturbative regime. The successful approaches are rather phenomenological, and contain  many internal parameters, which are fixed to reproduce the experimental data  (for reviews see Refs. \cite{Perdrisat:2006hj,Arrington:2006zm,Arrington:2011dn}).

On the other hand, the existing  elastic polarized and unpolarized $e^-p$ and  $e^+$p scattering  data  cover  kinematical region  broad  enough to reconstruct the FFs  dependence on $Q^2$. To combine the cross section data with the PT measurements and the $e^-p/e^+$p ratio data allows one to obtain  information about the TPE contribution.

The aim of this paper is to find the approximation of the FFs and the TPE contribution by mainly relying  on the experimental data. We  reduce the  model-dependent assumptions to the necessary minimum.
     \begin{figure}
     \centering{
     \includegraphics[width=9cm, height=7cm]{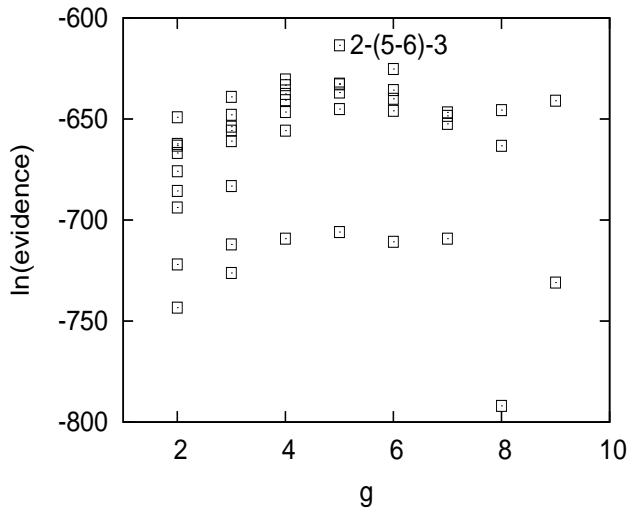}
     \caption{ Logarithm of evidence, see Eq. (\ref{log_of_evidence}). Each single point, in the plot, represents the ln(evidence) obtained for one particular NN architecture.   \label{Fig_evidence}}
     }
     \end{figure}

   \section{TPE and Neural Networks}

Only three complex FFs, which depend on $Q^2$ and $\epsilon$, are required to describe  the  elastic unpolarized and polarized $ep$ \cite{Guichon:2003qm} scattering amplitudes. Hence,  six real functions have to be determined from the data.

We assume that the PT ratio data are not affected by the TPE effect. Then, one can show that only three unknown functions have to be found: two proton FFs and the $\Delta C_{2\gamma}$ correcting term [see Eq. (\ref{sigma_r_2gamma})]. Analyses with similar TPE assumptions have been performed by many groups \cite{Chen:2007ac,Rekalo:2003xa,Borisyuk:2007re,TomasiGustafsson:2004ms,Arrington:2004ae,Arrington:2007ux,Alberico:2008sz,Arrington:2003df}.

 In this paper, we consider   the cross section, the PT, and the $e^+p/e^-p$ ratio data.  To consider  at least three different data types appeared to be necessary because of the limited model assumptions about the TPE term.

 To approximate the FFs and TPE function one has to assume particular empirical parametrization. However, it is obvious that the choice of the functional form of the parametrization  has an impact on the fit and its uncertainties. In particular, it is the case   of the  TPE contribution. This problem was not  discussed in the previous analyses.

 In the approach presented in this paper, fitting the data  means the construction of the statistical model with the ability to predict  the FFs and the TPE term.  We apply the methods of the Bayesian statistics, which allows  performing a model comparison. Indeed, we consider as many  different data parametrizations as possible, and the best model  is indicated by the objective Bayesian  procedure.

 In practice, one has to  evaluate the probability distribution $\mathcal{P}(model)$ in the space of all  functional parametrizations of the FFs and the TPE contribution. The best model  should maximize  this probability.
        \begin{figure}
        \centering{
\includegraphics[width=0.48\textwidth, height=12cm]{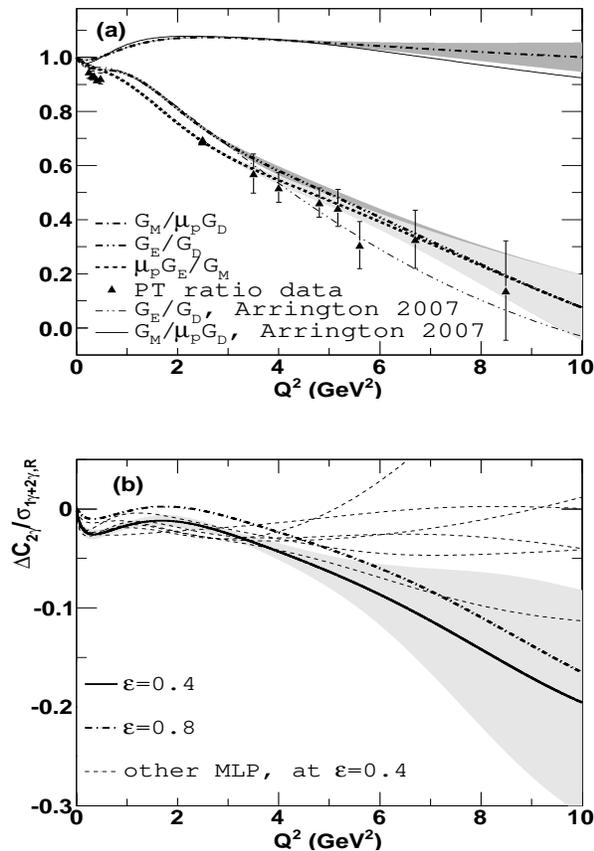}
\caption{(a) (Top panel) $G_E/G_D$, $G_M/\mu_p G_D$ [$G_D = 1/ (1 + Q^2/0.71~\mathrm{GeV}^2)^2$] and ratio $\mu_p G_E/G_M$. The predictions of the proton FFs of Arrington \textit{et al.} \cite{Arrington:2007ux} also are shown. The PT $\mu_p G_E/G_M$  data are taken from Refs. \cite{Ron:2011rdand_Puckett:2011xg,Puckett:2010ac,Meziane:2010xc}. Shaded areas denote  $1\sigma$ uncertainty.
 (b) (Bottom panel) The $Q^2$ dependence of  $\Delta C_{2\gamma}/\sigma_{1\gamma+2\gamma,R}$ at  $\epsilon=$0.4 and 0.8. The shaded area denotes $1\sigma$ uncertainty computed for the fit at $\epsilon=0.4$. The dotted lines  denote the TPE term predicted at $\epsilon=$ 0.4  by the networks that have lower than  best-fit evidence values.
 \label{Fig_FF_and_RATIO} }
}
\end{figure}

It is obvious  that the magnetic and electric FFs as well as the TPE correction function  are correlated. All of them should be determined by the same underlying fundamental model. Therefore, one can imagine that there exists a multidimensional function, defined by the set of parameters, which simultaneously describe all $G_M$, $G_E$ and $\Delta C_{2\gamma}$. In this paper, we use the artificial neural networks (ANNs) to approximate this function.   We consider a particular type of  ANN, the feed forward neural network (NN)  in the so-called multi layer perceptron (MLP) configuration.

The experimental data, which are analyzed here, depend on either one (only $Q^2$) or two ($Q^2$ and $\epsilon$)  kinematical variables.  Hence,
the MLP must map two-dimensional  input space [$\vec{\mathrm{in}} = (Q^2,\epsilon)^T$] to output space, spanned by three functions $\vec{\mathrm{out}} = (G_M, G_E, \Delta C_{2\gamma})^T$.

We consider MLP networks that consist of three layers of units: input, hidden layer, and output (see Fig. \ref{Fig_sieci}). Each single neuron (unit) of the network  calculates its  output value as an activation function $f_{act}$ of the weighted sum of its inputs $f_{act}\left(\sum_i w_i \mu_i\right)$, where $w_i$ denotes the i-th weight parameter, while $\mu_i$ represents the output value of the unit from the previous layer.

For the activation functions we take  the sigmoid $sigmoid(w)=1 / (1 + \exp(-w))$ and  the linear  functions for the hidden and output units, respectively. It has been proven \cite{Cybenko_Theorem} that the maps given by the networks with one hidden layer and with the sigmoid like activation functions, in this layer, are sufficient to approximate any continuous real function\footnote{The two-hidden layer NNs are sufficient to approximate any function.}.   Indeed, we assume that the FFs and TPE term are the continuous functions of kinematical variables. Notice that the efficiency of approximation depends on the number of  hidden units. In some problems it might be a very large number.

Additionally let us mention a useful property of the sigmoid function. Its effective support  is concentrated in the  close neighborhood of $w=0$. With increasing $|w|$, the sigmoid  saturates. This property allows  restricting the effective range of the weight parameters.

The $G_M$ and $G_E$ only depend  on $Q^2$. This property is achieved  by the particular choice of the architecture of MLP, namely some of the connections are erased. As a result,  the  network is divided into two sectors.  One, called the latter FF sector, which is disconnected with the  $\epsilon$ input and the second, called the latter TPE sector, which is connected with both input values.  The FFs and TPE correction are still  determined by the large subset of common weights.

An example of the 2-(3-2)-3  network  ($\mathcal{A}_{3,2}$) is drawn in Fig. \ref{Fig_sieci}. It consists of two input units, five units in the hidden layer (three units  belong to the FF sector, and two units belong to the TPE sector), and 3 output units.

The choice of the particular configuration of units and number of the hidden units, defines the network architecture $\mathcal{A}_{g,t}$. For given $\mathcal{A}_{g,t}$ the particular  map $\mathcal{N}_{g,t}$ is defined as
\begin{equation}
\mathcal{N}_{g,t}: \mathbb{R}^2 \mapsto \mathbb{R}^3,\quad \mathcal{N}_{g,t}(Q^2,\epsilon;\mathcal{A}_{g,t}, \vec{w}) = \pmatrix{G_M \cr G_E \cr \Delta C_{2\gamma}}.
  \end{equation}
To have  network  $\mathcal{A}_{g,t}$,  the optimal values of the weight parameters have to be found.   The process of establishing them is called the training of the network and it is described in the next part of the paper.

It is obvious that,  to find the optimal network architecture, which approximates desired map well, the number of hidden units has to be varied.  In this paper we apply the method that allows estimating the optimal size of the hidden layer.

  \section{Bayesian Framework}

The MLPs with a larger number of  units (with many weights) have a better ability to represent the data.  However, usually too complex parametrizations exactly resemble the data and usually tend to reflect the statistical fluctuations. Thus, the generality of the description is lost, and  the data are over-fitted. Moreover, the complex parametrizations may lead to larger uncertainties than the simple models. On the other hand, too simple parametrizations are not capable of coding  all the important information hidden in the measurements. The fits described by the simple functions might  be  characterized  by the underestimated  uncertainties.

A task of finding the optimal statistical model  that represents  the data accurately enough but does not overfit the data  is known in statistics as the bias-variance trade off problem. In the previous global analyses of the $ep$ data the degree of the complexity of the FFs and TPE parametrizations was chosen with the help of  phenomenological  arguments  and common sense.  In this paper, we wish to apply  the objective Bayesian methods, which allow quantitatively  investigating  the complexity of the FFs and TPE correcting functions.

The Bayesian framework (BF) formulated for the NN computations  \cite{bayes} faces the problems described above. This approach has already been  adapted to approximate the electromagnetic nucleon FFs \cite{Graczyk:2010gw} and, here, it is  developed to study the TPE effect. The BF was designed to:\footnote{A comprehensive introduction to Bayes' techniques in neural computation can be found in Ref. \cite{Bishop_book}.}
\begin{itemize}
      \item quantitatively classify the statistical hypothesis;
      \item objectively choose  the best network architectures and consequently, the number of hidden units;
      \item find the optimal values of the weight parameters;
      \item objectively establish  the training parameters, such as  the regularization parameter  $\alpha$ (it will be explained below).
      \end{itemize}

Notice that  to deal with the overfitting problem one can also use  the cross-validation technique. It  is complementary approach which has been applied by the NNPDF group for fitting the parton distribution functions \cite{Ball:2010de}.

At the beginning of  the Bayesian analysis, we assume that all possible models are equally likely,
\begin{equation}
\label{prior_assumtion}
\mathcal{P}(\mathcal{A}_{1,1})= ... =\mathcal{P}(\mathcal{A}_{g,t})=...,
\end{equation}
where $\mathcal{P}(\mathcal{A}_{g,t})$ denotes the prior probability.

With the help of Bayes' theorem  the posterior probability for a given model (network)   is obtained as
\begin{equation}
  \mathcal{P}(\mathcal{A}_{g,t}|\mathcal{D}) = \frac{\mathcal{P}(\mathcal{D}|\mathcal{A}_{g,t}) \mathcal{P}(\mathcal{A}_{g,t})}{\mathcal{P}(\mathcal{D})},
\end{equation}
where $\mathcal{D}$ is the experimental data,  $\mathcal{P}(\mathcal{A}_i|\mathcal{D})$ is the probability of the model given data $\mathcal{D}$, and $\mathcal{P}(\mathcal{D})$ is some constant real number. Because of the prior assumption (\ref{prior_assumtion}), it is obvious that,  to classify the
hypothesis,  it is enough to evaluate the evidence $\mathcal{P}(\mathcal{D}| \mathcal{A}_{g,t})$--the probabilistic measure of goodness of fit.

For  given network architecture $\mathcal{A}_{g,t}$,  the optimal $\vec{w}_{MP}$ weight parameters should maximize the posterior probability,
\begin{equation}
\label{posterior_w}
\mathcal{P}\left(\vec{w}\right|\left.  \mathcal{D}, \{\mathcal{I} \} ,\mathcal{A}_{g,t} \right) =
\frac{\mathcal{P}\left(\mathcal{D}\right|\left.\vec{w}, \{\mathcal{I} \}, \mathcal{A}_{g,t} \right)
\mathcal{P}\left(\vec{w}\right|\left. \{\mathcal{I} \}, \mathcal{A}_{g,t} \right)}{\mathcal{P}\left(\mathcal{D}\right|\left. \{\mathcal{I} \}, \mathcal{A}_{g,t} \right)},
\end{equation}
where $\mathcal{P}\left(\mathcal{D}\right|\left.\vec{w}, \{\mathcal{I} \}, \mathcal{A}_{g,t} \right)$ is the likelihood function of the data,  $\mathcal{P}\left(\vec{w}\right|\left. \{\mathcal{I} \}; \mathcal{A}_{g,t} \right)$ denotes the prior probability, and $\{I\}$ denotes the set of initial constraints.

The data likelihood function is defined by
\begin{equation}
\label{likelihood_function}
\mathcal{P}(\mathcal{D}|\vec{w}, \{\mathcal{I} \}, \mathcal{A}_{g,t} ) \sim \exp\left( - S_{ex}(\mathcal{D}, \vec{w}) \right),
\end{equation}
where
\begin{equation}
S_{ex}(\mathcal{D}, \vec{w}) = \chi^2_{\sigma} + \chi^2_{PT} + \chi^2_{\pm} + \chi_{G_M}^2+ \chi_{G_E}^2
\end{equation}
is the experimental error function. By $\chi^2_{\sigma, PT, \pm}$, we denote the error functions of the cross section (\ref{chi2_sigma}), the PT (\ref{chi2_PT})  and the $e^+p/e^- p $ ratio (\ref{chi2_positron}) data. Eventually, $\chi_{G_{M/E}}^2$ denotes the error function introduced to take  the two artificial FF points into account (see the discussion below).

We distinguish between  the ANN and  the physical initial constraints $\{ \mathcal{I} \} = \{ \mathcal{I} \}_{ANN}\cup \{ \mathcal{I} \}_{phys.}$.

The ANN constraints $\{\mathcal{I} \}_{ANN}$ are introduced  to   face the overfitting problem. Indeed, defining the prior probability as  follows:
\begin{eqnarray}
\label{prior_w}
\mathcal{P}\left(\vec{w}\right|\left. \{\mathcal{I} \}_{ANN}, \mathcal{A}_{g,t} \right) &=&
\mathcal{P}\left(\vec{w}\right|\left. \alpha, \mathcal{A}_{g,t} \right)
 \sim  \exp\left( - \alpha E_w(\vec{w}) \right) \nonumber \\ \\
 \label{penalty_term}
E_w (\vec{w}) & = & \frac{1}{2} \sum_{i\in all\, weights} w^2_i
 \end{eqnarray}
 prevents  getting the overfitted parametrizations.

The physical constraints $\{ \mathcal{I} \}_{phys.}$ are motivated by the general properties of the FFs and the TPE term \cite{Rekalo:2003xa,Chen:2007ac}. We assume that,  at $Q^2=0$, $G_M/\mu_p = G_E =1$ and $\Delta C_{2\gamma}( \epsilon=1)=0$. In practice,  three artificial data points are added to the experimental data sets, namely, [$G_M(0)/\mu_p =1, \Delta G_M(0)=\Delta $], [$G_E(0) =1, \Delta G_M(0)=\Delta $], and [$\mathcal{R}_{\pm}(0,1) = 1, \Delta \mathcal{R}_{\pm}(0,1) =\Delta$], where $\Delta =0.01$.   The influence of the $\Delta$ value on the fits and the training process were investigated in the preliminary stage of the analysis. It was obtained that, with $\Delta < 0.01$, the efficiency of the training process was very low, while retaining $\Delta > 0.01$ was not sufficient to attract the fit  for the desired value at the constraint points.

One can show that the maximum of the posterior probability (\ref{posterior_w}) corresponds to the minimum of the total error function,
\begin{equation}
\label{total_error_function}
S_{ex}(\mathcal{D},\vec{w}) + \alpha E_w(\vec{w}).
\end{equation}
Let $\vec{w}_{MP}$ denote the weight configuration, which  minimizes the above  expression. To find the minimum (\ref{total_error_function}), the quick-prop gradient descent algorithm \cite{quick-prop}, is applied. The weight parameters are  updated iteratively, because of the algorithm, until the minimum is reached.

The proper choice of the $\alpha$ parameter is crucial for getting the fits and for further model comparison. If the $\alpha$ parameter is small then the penalty term (\ref{penalty_term}) does  not significantly affect  the results of the training process. In the language of the Bayesian statistics, a small $\alpha$ value corresponds to the  large width of the  prior probability distribution (\ref{prior_w}).

The BF provides a recipe on how to establish the optimal  $\alpha$ parameter. For the optimal $\alpha_{MP}$ and $\vec{w}_{MP}$  the probability distribution,
\begin{equation}
\mathcal{P}\left(\alpha\right|\left. \mathcal{D}, \mathcal{A}_{g,t} \right)  = \frac{\mathcal{P}\left(\mathcal{D}\right|\left. \alpha, \mathcal{A}_{g,t} \right) \mathcal{P}\left(\alpha\right|\left.  \mathcal{A}_{g,t} \right)}{\mathcal{P}\left(\mathcal{D}\right|\left.\mathcal{A}_{g,t} \right)}
\end{equation}
is maximized.

From the above expression, one can obtain, the  necessary condition (\ref{alpha_MP_parameter}), which has to be satisfied by $\alpha_{MP}$. In this version of the BF approach, we consider the so-called ladder approximation, where the expression (\ref{alpha_MP_parameter}) is used to iteratively update   the $\alpha$ value (during the training process), as long as it converges. In reality, because  Eq. (\ref{alpha_MP_parameter})  is valid in the neighborhood of the minimum, the $\alpha$ parameter is not changed until  the training process  approaches  the  close surrounding of the minimum. In this part of the training $\alpha_0$ is fixed and equals 0.01. Then, it  starts to be updated.

In general, for every weight parameter, one would introduce  its own regularization factor. However, notice that $\alpha$ is an example of the scale parameter of the model,  given that network $\mathcal{A}_{g,t}$ is  symmetric under the permutation of units in the hidden layer.\footnote{Appropriate permutation  of he units  in the hidden  layer of either the FF and/or the  TPE sectors does not change the output response.}.  This property  allows reducing the number of independent regularization parameters to six: three in the FF sector (hidden, bias, and linear regularization factors) and three in the TPE sector (similar to before). On the other hand,  $\Delta C_{2\gamma}$ is given by the linear combination of the nucleon FFs, multiplied by the additional TPE-like FFs (see Eq. (14) of Ref. \cite{Carlson:2007sp}). It means that, if the parameters of the FF sector are scaled, then the weights of the TPE sector should also be  rescaled. This property only seems to be  approximate, but we use it to reduce the number of regularization parameters to three. Eventually, we noticed that in our previous paper, it was shown that it was enough to consider one regularization parameter  to fit the FF data \cite{Graczyk:2010gw}.  Hence,   to  simplify the numerical calculations and also to accelerate the training process (more than 45 000 training processes have been performed),  we consider the simplest regularization scenario with one $\alpha$ parameter. However, as  described above, this part of the approach can be improved \cite{future}.

By having the optimal values of the weight and $\alpha$ parameters the evidence is computed from Eq. (\ref{log_of_evidence})   The logarithm of evidence  is given by two main contributions: the misfit of the approximate data (the experimental error function at the minimum) and the Occam factor. The latter  penalizes  complex models. The  most optimal model  has the  highest value of evidence. The evidence formula and the description of its  properties can be found in Appendix \ref{Appendix_Evidence}.

\begin{figure}
\centering{
\includegraphics[width=0.48\textwidth, height=12cm]{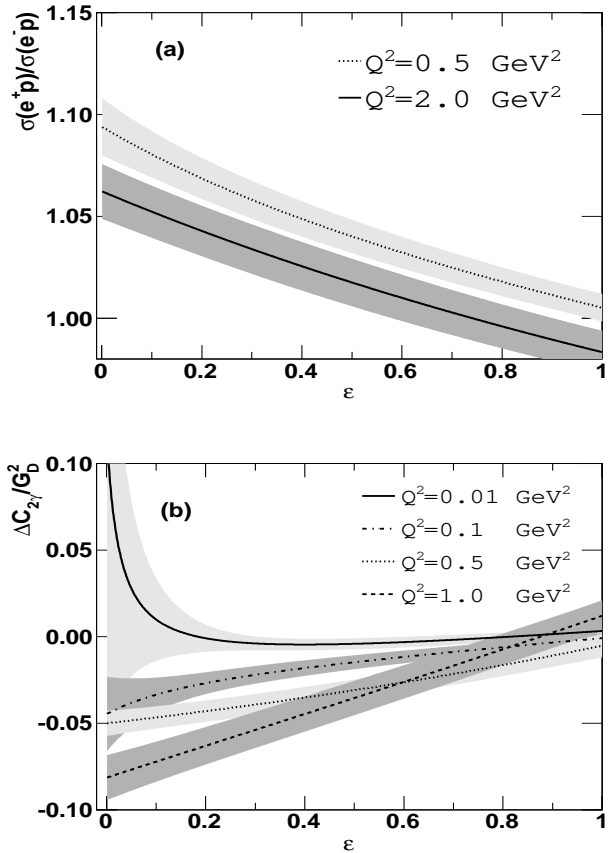}
\caption{(a) (Top panel) Ratio $\mathcal{R}_{\pm}$ predicted by the network 2-(5-6)-3. The grey areas denote $1\sigma$ uncertainty.
         (b) (Bottom panel) $\Delta C_{2\gamma}/G_D^2$ dependence on $\epsilon$. The darker and lighter gray areas denote $1\sigma$ uncertainty. \label{Fig_Ratio_POS} }
}
\end{figure}

  \section{Numerical Analysis}
  \label{section_numerical_analysis}

As  mentioned in the previous section, we consider  three types of  measurements: the cross section (27 sets), the PT (14 sets)  and the $e^+p/e^- p $ ratio (3 sets) data.

The selection of the cross section and PT ratio data sets is the same as in one of our previous papers \cite{Alberico:2008sz}. However, in the case of the PT data,  two data sets are replaced with their recent updates \cite{Ron:2011rdand_Puckett:2011xg}. Additionally, we also include  the latest PT measurements of the FF ratio \cite{Meziane:2010xc,Puckett:2010ac}.  Since the presence of the PT ratio  data is required to properly extract the TPE contribution, we only consider  the  cross section points  below $Q^2=10$~GeV$^2$. Above this limit, the PT data are not available.

In the case of the  cross section data, similar to Ref. \cite{Alberico:2008sz}, the systematic normalization uncertainties are taken into account. For every data set, a normalization parameter is introduced and it is established during  training. The procedure is described in Appendix \ref{Appendix_error_normalization}.

We  consider  the networks of type 2-(g-t)-3, where $ 4 \leq g+t = M \leq 12$. In the preliminary stage of the analysis, it has been observed that the networks with either g=1, or t=1 have not been  able to approximate the data well (similar to the networks with  $M < 4$). Therefore we only consider  models with $g,t >1 $. Finally we discuss 45 different ANN architectures.  For every  network $\mathcal{A}_{g,t}$ type   $10^3$ networks, with  randomly chosen initial values of weights, have been trained. Among them, the parametrization with the highest evidence  was used for  further model comparison.  It turned out that    the highest evidence value was obtained for  network 2-(5-6)-3 (see Fig. \ref{Fig_evidence}).

In Fig.~\ref{Fig_FF_and_RATIO}(a),  we plot  the FF ratio $\mathcal{R}_{1\gamma}$ computed with
 the  network $\mathcal{A}_{5,6}$ (our best fit).  The shaded areas denote $1\sigma$ uncertainty computed from the covariance matrix of the fit.
 Our predictions of the FFs are compared to the results of Ref. \cite{Arrington:2007ux} where the TPE function was postulated based on the  phenomenological  arguments. The  discrepancy between our fits and those of Ref.
 \cite{Arrington:2007ux}  appears above $Q^2$=4 GeV$^2$.

 In  Fig.~\ref{Fig_FF_and_RATIO}(b), the $Q^2$ dependence of ratio $\Delta C_{2\gamma}/\sigma_{1\gamma+2\gamma}$ is presented. We see that at $Q^2 \sim 0.2$~GeV$^2$ the TPE correcting term has a local minimum and  it becomes the decreasing function of $Q^2$ above 2~GeV$^2$. With growing $Q^2$, fit uncertainty also enlarges.  Indeed, above $Q^2=6$~GeV$^2$, the number of experimental points is limited and the data are not accurate enough to get an exact approximation. It is interesting to mention  that, for large $\epsilon$ (above 0.8) and   $Q^2$  around 1.5~GeV$^2$,  the TPE correction  is positive.

 In  Fig.~\ref{Fig_FF_and_RATIO}(b), we also plot  the TPE contribution (dotted lines) predicted by the models: $\mathcal{A}_{4,2}$, $\mathcal{A}_{4,3}$, $\mathcal{A}_{6,2}$, $\mathcal{A}_{6,3}$, $\mathcal{A}_{6,4}$, and $\mathcal{A}_{5,7}$. They are characterized by lower evidence values than the $\mathcal{A}_{5,6}$ model, but they could be acceptable because of the $\chi^2$ method (their $\chi^2_{min}$ values are much lower than the number of points). The difference between these fits and the prediction of $\mathcal{A}_{5,6}$ is spectacular. It demonstrates that the model comparison is crucial for the proper choice  of  TPE parametrization.

By keeping  the forthcoming measurements of the elastic $e^-p$ and $e^+ p$ scatterings  \cite{positron_new} in mind, in Fig. \ref{Fig_Ratio_POS}(a), we plot our predictions of ratio $\mathcal{R}_{\pm}$. Although, we have not assumed the linearity  of the TPE term  in $\epsilon$, the final fit  behaves like a linear function of  $\epsilon$, as  observed in the previous global analysis \cite{Tvaskis:2005ex}. Although  nonlinearities appear at the low $\epsilon$ and $Q^2$ values (bottom panel of Fig. \ref{Fig_Ratio_POS}),  in this kinematical domain, the fits have large uncertainties, and the obtained results are in agreement with the linear approximation.

The obtained  TPE function has a particular analytical form (see Appendix \ref{Appendix_fits}), which can be written as the Taylor series in $\epsilon$. If one  neglects higher rather than linear $\epsilon$ terms, then the TPE correction is  the sum of two contributions, which play a particular role in the LT separation. One of them corrects  the magnetic FF, and it appears to be negative. The other  modifies the electric FF, and it is the positive function of $Q^2$.

Notice that the  PT data are not present below $Q^2=0.16$~GeV$^2$. Hence its influence on the extraction of the TPE in this kinematical range is small. Although because of the lack of PT data,  the TPE is still constrained in the low $Q^2$ region. Namely, there are   several $e^+p/e^- p$ ratio data points \cite{Yount62}. Additionally, we keep   one artificial point at $Q^2=0$, and $\epsilon=1$ which constrains $\Delta C_{2\gamma}$. Also,  there are  plenty of  cross section data points  and  the two artificial FF points. The presence of the  FF points determines the low $Q^2$ behavior of the $\sigma_{1\gamma,R}$.  All together, provides restrictions  on the extraction of the TPE term.

In  Fig. \ref{Fig_Ratio_POS}(b), we show the $\epsilon$ dependence of $\Delta C_{2\gamma}$ at several values of $Q^2$. It can be seen  that at very low $Q^2$, the TPE term becomes positive. However, similar to the above, because of the large uncertainties, this effect is consistent with $\Delta C_{2\gamma}=0$.

The aim of this paper was to find the approximation of  the proton FFs and the TPE function based on the knowledge of the elastic $ep$ scattering data.  It was performed by adapting the Bayesian statistical methods developed for the feed forward NNs. We assumed that the TPE correction does not affect the PT ratio data. This assumption turned out to be necessary to perform the numerical analysis, but one should keep in mind that there  is no perfect approximation at low $Q^2$.

We discussed as many different NN parametrizations as  required to find the optimal model. The best model was indicated by the Bayesian algorithm. From this point of view the results are model independent. The obtained TPE fit turned out to have nontrivial $Q^2$ dependence. In some kinematical regions (very low $Q^2$ and $Q^2\sim 1.5$~GeV$^2$, $\epsilon > 0.8$), it is the positive function.

Let us emphasize that  we considered the simplest working BF. Only one regularization parameter was discussed, and the Hessian approximation was  applied. Further development of the approach might improve the results of the analysis. In particular, it allows going beyond the covariance matrix approximation used for the estimation of the fit uncertainty. The improvements require introducing  modifications at  every step of the BF.  The new approach will also need  greater  computational power than the present one.

The analytical form of the fits is shown in Appendix \ref{Appendix_fits}, whereas the covariance matrix can be taken  from Ref. \cite{graczyk_web}. All numerical computations have been performed with the C++ library developed by K.M.G.

\section*{Acknowledgements}

This work was supported by the Polish Ministry of Science Grant,  Project No.	N N202 368439.
We thank C. Giunti for  inspiring discussions in the early  stage of the project. We acknowledge   very instructive discussions with R. Sulej and P. Plonski, we thank  J. Zmuda and J. Nowak for reading the manuscript and we thank  J. Arrington for his remarks  on the previous version of the paper.

\appendix

\section{Error Functions}
\label{Appendix_error}
The cross section error function reads as
\begin{equation}
\label{chi2_sigma}
\chi^2_\sigma = \sum_{k=1}^{N_\sigma} \left[ \sum_{i=1}^{n_k} \left(\frac{\eta_k \sigma^{th}_{ki} - \sigma^{ex}_{ki}}{\Delta\sigma_{ki}}\right)^2 +
\left(\frac{\eta_{k}-1}{\Delta \eta_k } \right)^2 \right],
\end{equation}
where $N_\sigma$ is the number of independent cross section data sets,
$n_k$ is the number of points in the $k$th data set,
$\eta_k$ is the normalization parameter for the $k$th data set, $\Delta \eta_k$ is the normalization (systematic)  uncertainty,
$\sigma^{ex}_{ki}$ is the experimental value of the reduced cross section of  the $i$th data point in the $k$th data set measured for  $Q_{ki}^2$ and $\epsilon_{ki}$, $\Delta\sigma^{ex}_{ki}$ denotes the corresponding experimental uncertainty, and
$\sigma^{th}_{ki} \equiv  \sigma_{1\gamma+2\gamma, R}(Q_{ki}^2, \epsilon_{ki})$.

The PT ratio data error function reads as
\begin{equation}
\label{chi2_PT}
\chi^2_{PT} =  \sum_{i=1}^{n_k^{PT}} \left(\frac{ \mathcal{R}^{th}_{i} - \mathcal{R}^{ex}_{i}}{\Delta\mathcal{R}_{i}}\right)^2,
\end{equation}
where $n_k^{PT}$ is the number of PT ratio data points, $\mathcal{R}^{ex}_{i}$  denotes the experimental value of the $i$th point, measured for $Q^2_i$,
$\Delta \mathcal{R}^{ex}_{i}$ is the corresponding experimental uncertainty, and
$\mathcal{R}^{th}_{i} \equiv \mathcal{R}_{1\gamma}(Q^2_i)$.

Analogically the positron-electron ratio data error function reads as
\begin{equation}
\label{chi2_positron}
\chi^2_{\pm} =  \sum_{i=1}^{n_k^{\pm}} \left(\frac{ \mathcal{R}^{\pm,th}_{i} - \mathcal{R}^{\pm,ex}_{i}}{\Delta\mathcal{R}_{i}^{\pm}}\right)^2,
\end{equation}
where $n_k^{\pm}$ is the number of PT ratio data points, $\mathcal{R}^{\pm,ex}_{i}$  denotes the experimental value of the $i$th point, measured for $Q^2_i$ and $\epsilon_i$ values,  $\Delta \mathcal{R}^{\pm,ex}_{i}$ is the corresponding experimental uncertainty, and $\mathcal{R}^{\pm,th}_{i} = \mathcal{R}_{\pm}(Q^2_i, \epsilon_i)$.

The $\chi^2_{G_M/G_E}$ reads as
\begin{equation}
\label{chi2_G}
\chi^2_{G} = \left(\frac{G-1}{\Delta}\right)^2,
\end{equation}
where
$G=G_M/\mu_p$ or $G_E$.

\section{Data Normalization}
\label{Appendix_error_normalization}
The optimal values of the normalization parameters $\eta_k$, $k=1,2,...,N_\sigma$, at the minimum of the total error function (\ref{total_error_function}), must satisfy the property,
\begin{equation}
0 = \frac{\partial S_{ex}}{\partial \eta_k}, k=1,...,N_\sigma,
\end{equation}
which can be rewritten as
\begin{equation}
\label{normalization_condition}
\eta_k = \frac{ \displaystyle \sum_{i=1}^{n_k}
\frac{\sigma^{th}_{ki} \sigma^{ex}_{ki}}{(\Delta \sigma_{ki})^2}
+
\frac{1}{(\Delta \eta_k)^2}}{\displaystyle \sum_{i=1}^{n_k} \frac{(\sigma^{th}_{ki})^2 }{(\Delta \sigma_{ki})^2} + \frac{1}{(\Delta \eta_k)^2}}.
\end{equation}
The above expression is used for updating the normalization parameters during training.  The procedure turned out to be convergent, as long as the minimum of the total error function was reached. It is interesting to notice that the normalization parameters obtained in this paper are very similar to those obtained in our previous global analysis  \cite{Alberico:2008sz}, where  the MINUIT package (now it is  one of the packages of the root library) was applied to find the optimal values of the fit and normalization parameters.

\section{Regularization Parameter}
\label{Appendix_alpha}

The $\alpha_{MP}$ parameter  is computed in the so-called Hessian approximation \cite{Bishop_book}. It is given by the solution of the equation,
\begin{equation}
\label{alpha_MP_parameter}
 2 \alpha_{MP}E_w(\vec{w}_{MP}) =    \sum_{i=1}^W \frac{\lambda_i}{\lambda_i + \alpha_{MP}} \equiv \gamma,
\end{equation}
where $\lambda_i$'s are eigenvalues of the  matrix $ \nabla_n \nabla_m \left. S_{ex} \right|_{\vec{w} = \vec{w}_{MP}}$ and $\nabla_i \equiv \partial_{w_i}$.

In practice, to find the optimal  $\alpha_{MP}$, the  $\alpha$ parameter is iteratively updated during the training process, i.e.,
\begin{equation}
\label{alpha_update_rule}
\alpha_{k+1} = \gamma(\alpha_{k})/2E_w(\vec{w}),
\end{equation}
where $\alpha_k$ denotes the value of the normalization parameter in the $k$th iteration step of training.

\section{Evidence}
\label{Appendix_Evidence}

The logarithm  of evidence, the terms that are the same for different network architectures that are omitted, reads as
\begin{widetext}
\begin{equation}
\label{log_of_evidence}
\ln \mathcal{P}\left(\mathcal{D}\right|\left. \mathcal{A}_{g,t} \right)
 \approx  \underbrace{-S_{ex}(\mathcal{D},\vec{w}_{MP})}_{misfit} \underbrace{- \alpha_{MP} E_{w}(\vec{w}_{MP})
-\frac{1}{2}\ln |A| + \frac{W}{2}\ln \alpha_{MP} -\frac{1}{2}\ln \frac{\gamma}{2} + \overbrace{(g+t) \ln(2) + \ln(g!)+\ln(t!)}^{symmetry\;factor}}_{Occam\; factor},
\end{equation}
\end{widetext}
where $W$ is the number of weigh parameters, and $|A|$ is the determinant of the Hessian  matrix $A_{ij} = \nabla_i \nabla_j \left. S_{ex} \right|_{\vec{w}=\vec{w}_{MP}} + \alpha_{MP}$.

The first term  in Eq. (\ref{log_of_evidence}), usually of low-value, is the misfit of the approximated data, while the other terms
contribute to the Occam factor. The latter penalizes the complex models. An additional contribution to this quantity is given by the symmetry factor.

In the network  $\mathcal{A}_{g,t}$, some hidden units can be interchanged, but the response of the network (output values) reminds unchanged. It means that for a given configuration of weights, there exists some number of equivalent networks, which differ only by the appropriate permutation of the weights. It gives rise to the appearance of the additional combinatorial factor in the evidence. However, in this paper, it does not play a significant role.

\section{Analytical Form of  Fits}
\label{Appendix_fits}
We now have,
\begin{eqnarray}
\frac{G_M (Q^2)}{\mu_p G_D(Q^2)} &=& \sum_{i=3}^{7} w_{i,15} f_{act}( Q^2 w_{0,i} + w_{2,i}) + w_{14,15}, \nonumber \\
\\
\frac{G_E (Q^2)}{ G_D(Q^2)} &=& \sum_{i=3}^{7} w_{i,16} f_{act}( Q^2 w_{0,i} + w_{2,i}) + w_{14,16}, \nonumber \\ \\
\frac{ \Delta C_{2\gamma} (Q^2, \epsilon)}{ G_D^2(Q^2)} &=& \sum_{i=3}^{13} w_{i,17} f_{act}( Q^2 w_{0,i}+ \epsilon w_{1,i} + w_{2,i}) + w_{14,17}, \nonumber \\
\\
G_D (Q^2) &=& \frac{1}{\left(1 + Q^2/0.71 \right)^2}, \\
f_{act}(x)&=& \frac{1}{ 1 + \exp(-x)},
\end{eqnarray}
\begin{scriptsize}
\begin{eqnarray*}
w_{3,17} &=&-0.7994, \quad w_{4,17} = 0.9775, \\
w_{5,17} &=& 4.70641,  \quad w_{6,17} =-0.5378, \\
w_{7,17} &=&0.20026,  \quad w_{8,17} =0.08842,  \\
w_{9,17} &=&-5.25238, \quad w_{10,17} = 6.91219,  \\
w_{11,17} &=&-4.09499, \quad w_{12,17} =1.53302,  \\
w_{13,17} &=&-1.29911,  \quad w_{14,17} =-2.32656,  \\
w_{3,16}& =&  0.87769,  \quad w_{4,16}  = -1.42417,  \\
w_{5,16}& =&  5.31229,  \quad w_{6,16}  = -7.03220,  \\
w_{7,16}& =&  1.16534,  \quad w_{14,16} = 1.64949,  \\
w_{3,15}& =&  0.06206,  \quad w_{4,15} = -0.16708,  \\
w_{5,15}& =& -2.05062,  \quad w_{6,15} = -3.18562,  \\
w_{7,15}& =&  1.43697,  \quad w_{14,15} = 4.91944  \\
w_{0,13}& =&  0.01442,  \quad w_{0,12} =  0.14829, \\
w_{1,13}& =&  0.15599,  \quad w_{1,12} =  0.50796,  \\
w_{2,13}& =&  0.34353  \quad w_{2,12}  = -0.91625,  \\
w_{0,11}& =& 0.41505,  \quad w_{0,10}  =-0.44004,  \\
w_{1,11}& =&-0.13263,  \quad w_{1,10} = 0.65672,  \\
w_{2,11}& =& 1.80434,  \quad w_{2,10} = 3.66358,  \\
w_{0,9}& =&  0.37908,  \quad w_{0,8} =  0.01100, \\
w_{1,9}& =&  0.33989,  \quad w_{1,8} =  0.05041, \\
w_{2,9}& =&  1.25938,  \quad w_{2,8} =  0.68383, \\
w_{0,7}& =&  1.02290,  \quad w_{0,6} =  3.26523,  \\
w_{2,7}& =&  0.97734,  \quad w_{2,6} =  3.63669,  \\
w_{0,5}& =&  0.71814,  \quad w_{0,4} =  0.25270,  \\
w_{2,5}& =&  2.42620,  \quad w_{2,4} = -1.60333,  \\
w_{0,3}& =& -1.22837  \quad w_{2,3} =  1.44729, \\
w_{1,3}&=& w_{1,4}= w_{1,5}= w_{1,6}= w_{1,7}=0.
\end{eqnarray*}
\end{scriptsize}

Notice that $Q^2$ in the above formulas is meant to be in units of GeV$^2$.

The FF and TPE parametrizations are obtained for $Q^2 \in (0, 10~\mathrm{GeV}^2) $ and $\epsilon \in (0,1)$. However, one should keep  the problems of the extraction of TPE at very low $Q^2$ in mind (see the discussion in the last section of the paper).

\end{document}